\documentclass[12pt,aps,prb,article]{revtex4}
\begin{document}
\title{Comment on `Frozen time in hyperbolic spacetime motion'}
\author{Jerrold Franklin}
\email{Jerry.F@TEMPLE.EDU}
\affiliation{Department of Physics\\
Temple University, Philadelphia, PA 19122-6082}
\date{\today}
\begin{abstract}
We show that the conclusion in version 3 of the paper `Frozen time in hyperbolic spacetime motion' that time does not move in a spaceship undergoing hyperbolic motion is wrong because of a trivial error in interpreting the distance and time variables used in the paper, and because it now incorrectly states that ``Hyperbolic motion does not imply constant acceleration."
\end{abstract}
\maketitle

This Comment is on version 3\cite{b3} of the paper `Frozen time in hyperbolic spacetime motion'.  Version 1\cite{jf1} of our Comment on the arXiv showed that version 1\cite{b1} of the paper was wrong, and that Comment is still appropriate for that version.  
For completeness, we reproduce the first two paragraphs of version 1 of our Comment:

``The paper `Frozen time in hyperbolic spacetime motion' purports to show that time does not move in a spaceship undergoing hyperbolic motion resulting from a
constant acceleration in its rest system.   B thus comes to the absurd conclusion that a spaceship can `reach distant worlds without the crew aging at all'.
The derivation in  B1 and the conclusion reached are completely wrong.  

The error in B1 lies in misinterpreting  the space and time coordinates in equation (6) of the paper
\begin{equation}
x^2-c^2t^2 = x'^2-c^2t'^2,
\end{equation}
where $x$ and $t$ are the space and time coordinates of the spaceship in its original rest frame S, and $x'$ and $t'$ its space and time coordinates in a Lorentz system S$'$ moving at a constant velocity $v$ such that the spaceship is momentarily at rest in system S$'$.  B claims that since the spaceship does not move in its rest system, the coordinate $x'$ in Eq. (1) is fixed at $x_0$, its initial position when both $t$ and $'t$ are zero.  B then concludes that, since $x'$ is fixed at its initial value, $t'$ must also remain fixed at its initial value of 0.  However, the coordinate $x'$ in Eq.  (1) is the coordinate of the space ship in the system S$'$ which only momentarily has the same speed as the spaceship.  In system S$'$ the spaceship starts out moving with velocity $-v$ and then decelerates coming momentarily at rest in system S$'$.  Consequently, the spaceship moves an appreciable distance in system S$'$, and $x'$ is not at all 0.  This fact destroys the reasoning in B."

The above comment still applies to B3, because it still misinterprets the tine variables $x'$ and $t'$ in its Eqs. (6) and (7), but it compounds the confusion with more errors.
The new version B3 retracts the use in B1 of constant acceleration in the instantaneous rest frame, because critics had pointed out that numerous textbooks have shown that constant acceleration in the instantaneous rest frame precludes `frozen time'.
However, B3 now makes the false assertion that ``Hyperbolic motion does not imply constant acceleration."  
On the basis of this, B3 disavows the use of constant acceleration.
This repudiates and makes most of B1 irrelevant, as well as the first paragraph of Section 2 of B3 which still uses constant acceleration.

Without constant acceleration, basic equation (7) of B3 no longer holds.  So, while Eq. (7) was correct in B1, but misinterpreted, now (without constant acceleration)
 it is not even correct, while still misinterpreted.
This, in fact, is shown in `Additional comment 1' of FAQ \#2a of B3.  For completeness, we repeat the derivation below (in the notation of B3):
``From (1) in Sect 2 of the preprint (B3) we have $x = \sqrt{\tau^2 + x_0^2}$,
and thus get the velocity $\beta=dx/d\tau=\tau/\sqrt{\tau^2+x_0^2}$,
and thus also $\gamma=1/\sqrt{1-\beta^2}=\sqrt{\tau^2+x_0^2}/x_0$.
Differentiating the velocity to get the acceleration, we then get after simplification 
$\alpha=\gamma^{-3}/x_0$,
which transformed to the co-moving frame becomes $a'=1/x_0$.
Since this is a constant, it would
seem that hyperbolic motion with necessity must be coupled to a constant acceleration in the comoving
frame."

This simple derivation, starting with the equation of hyperbolic motion [either Eq. (1) or Eq. (7) of B3] and finding only constant acceleration $a'$ in the comoving frame,
clearly shows that hyperbolic motion requires constant acceleration.
However B3 turns this on its head by following this derivation with the contradictory sentence
``However, whereas thus constant acceleration implies hyperbolic motion, the opposite
is not true: Hyperbolic motion does not imply constant acceleration."  Perhaps another reader can make sense of this.

The version B3 gives another derivation on pages 8 and 9 involving the hyperbolic functions sinh and cosh.
The full derivation can be seen in B3, but we summarize it here.
The derivation again shows that ``the momentary acceleration in the co-moving frame is thus always constant $1/x_0$."
But, just after showing that the acceleration must be constant, still insists ``We can have a hyperbolic motion as in (6) above regardless of whether
the acceleration is constant or not in the co-moving frame."

In summary, version 3 of `Frozen time in hyperbolic spacetime motion' has same error of interpretation as version 1, but adds more errors involving hyperbolic motion.
The conclusion in either version that ``a spaceship can reach distant worlds without the crew aging at all" is not only absurd, but is also wrong.

\end{document}